%% file: arxiv.tex
\documentclass[10pt,conference]{IEEEtran}
\usepackage{cite}
\usepackage{amsmath,amssymb,amsfonts}
\usepackage{algorithmic}
\usepackage{graphicx}
\usepackage{textcomp}
\usepackage{xcolor}
\usepackage[hyphens]{url}
\usepackage{fancyhdr}
\usepackage{hyperref}
\usepackage{booktabs}

\usepackage{multirow}
\usepackage{bm}
\usepackage{graphicx}

\usepackage{url}
\usepackage{subcaption}
\usepackage{subcaption}
\usepackage[percent]{overpic}
\usepackage{stmaryrd}
\usepackage{dcolumn}
\usepackage{amssymb}
\usepackage{placeins}
\usepackage{quantikz}
\usepackage[group-separator={,}]{siunitx}

\usepackage{braket}
\usepackage{amsmath}
\usepackage{afterpage}
\usepackage{xcolor}
\usepackage{verbatim}

\title{A Heterogeneous Distributed Architecture for Quantum Simulation}

\author{
\IEEEauthorblockN{
John Stack\textsuperscript{1,2,*},
Sitong Liu\textsuperscript{3,2},
Abhinav Anand\textsuperscript{2},
Inder Monga\textsuperscript{2},
Yuan Liu\textsuperscript{1,5,6},
\\[2pt]
Erhan Saglamyurek\textsuperscript{2},
Pedro L. S. Lopes\textsuperscript{4,*},
Frank Mueller\textsuperscript{1,*},
and Katherine Klymko\textsuperscript{2,*}
}

\IEEEauthorblockA{
\textsuperscript{1}
Department of Computer Science,
North Carolina State University,
Raleigh, North Carolina, USA
\\
\textsuperscript{2}
Lawrence Berkeley National Laboratory,
Berkeley, California, USA
\\
\textsuperscript{3}
Department of Electrical and Computer Engineering,
Duke University,
Durham, North Carolina, USA
\\
\textsuperscript{4}
QuEra Computing Inc.,
Boston, Massachusetts, USA
\\
\textsuperscript{5}
Department of Electrical and Computer Engineering,
North Carolina State University,
Raleigh, North Carolina, USA
\\
\textsuperscript{6}
Department of Physics and Astronomy,
North Carolina State University,
Raleigh, North Carolina, USA
\\[2pt]
\textsuperscript{*}Corresponding authors: jstack@ncsu.edu, plopes@quera.com, fmuelle@ncsu.edu, kklymko@lbl.gov
}
}

\begin{document}
\pagestyle{fancy}
\fancyhf{}
\fancyfoot[C]{\thepage}
\renewcommand{\headrulewidth}{0pt}
\renewcommand{\footrulewidth}{0pt}

\maketitle
\thispagestyle{fancy}

\begin{abstract}
\input abstract
\end{abstract}

\input intro
\input arch
\input related
\input eval

\input perform

\input storage

\input network

\input para
\input disc
\input methods

\section*{Acknowledgments}
We thank Yifan (Frank) Zhang for helpful comments on the manuscript. J.S. and S.L. were summer interns at Lawrence Berkeley National Laboratory during the initiation of this project. J.S. and F.M. acknowledge support from NSF grants OSI-2531350, OSI-2410675, PHY-2325080 and OMA-2120757 as well as DOE grant DE-SC0025384. S.L. acknowledges support from the National Science Foundation STAQ project under Grant No. PHY-2325080. This research used resources of the National Energy Research Scientific Computing Center, a DOE  Office of Science User Facility supported by the Office of Science of the U.S. Department of Energy under Contract No. DE-AC02-05CH11231 using NERSC award ASCR-ERCAP0037552. Y.L. acknowledges support from DOE grant DE-SC0025384. This work was also supported in part by the U.S. Department of Energy, Office of Science, under Award No. DE-SCL0000039
to Lawrence Berkeley National Laboratory (PI: E.S.). 

\section*{AI Use}
Claude Opus 4.8 and ChatGPT 5.5 were used to extend and run the simulation code in this paper. They were also used to generate code to produce the figures in this paper from the simulator data. The diagrams (Fig.\ref{fig:simulation_workflow} and Fig.\ref{fig:arch}) were made by a human. 

\bibliographystyle{IEEEtranS.bst}
\bibliography{apssamp}

\end{document}

%% file: abstract.tex
Architectural specialization and distribution can help scale fault-tolerant quantum computers, but may also introduce substantial overheads from communication, routing, and resource duplication. We introduce a heterogeneous distributed architecture in which a magic core is connected to an extensible storage system composed of one-dimensional lanes of specialized cold-storage nodes. The storage system supports parallel random access to Pauli string parities.

This organization is particularly well suited to fermionic quantum simulation, enabling parallel execution of the highly non-local Pauli strings arising from these systems. We evaluate the architecture on fault-tolerant simulations of the dynamics of the Fermi–Hubbard and sparse Sachdev–Ye–Kitaev (SYK) models on systems of up to 450 logical qubits. These workloads exhibit complementary communication structures: Fermi–Hubbard produces a spectrum of interactions from local to non-local shaped by lattice geometry, whereas sparse SYK produces highly non-local and overlapping Pauli operators. For a Trotter step of a 450-logical-qubit Fermi–Hubbard workload, a six-lane system with 30 T-state factories is within approximately 1.4× the wall-clock time of a homogeneous distributed architecture with 4 times as many T-state factories and substantially greater connectivity and sites for injecting magic. For matched T-factory counts, our architecture is $\sim 2\times$ faster.

%% file: intro.tex
\section{Introduction}

\begin{figure*}[!t]
  \centering
\includegraphics[width=\linewidth]{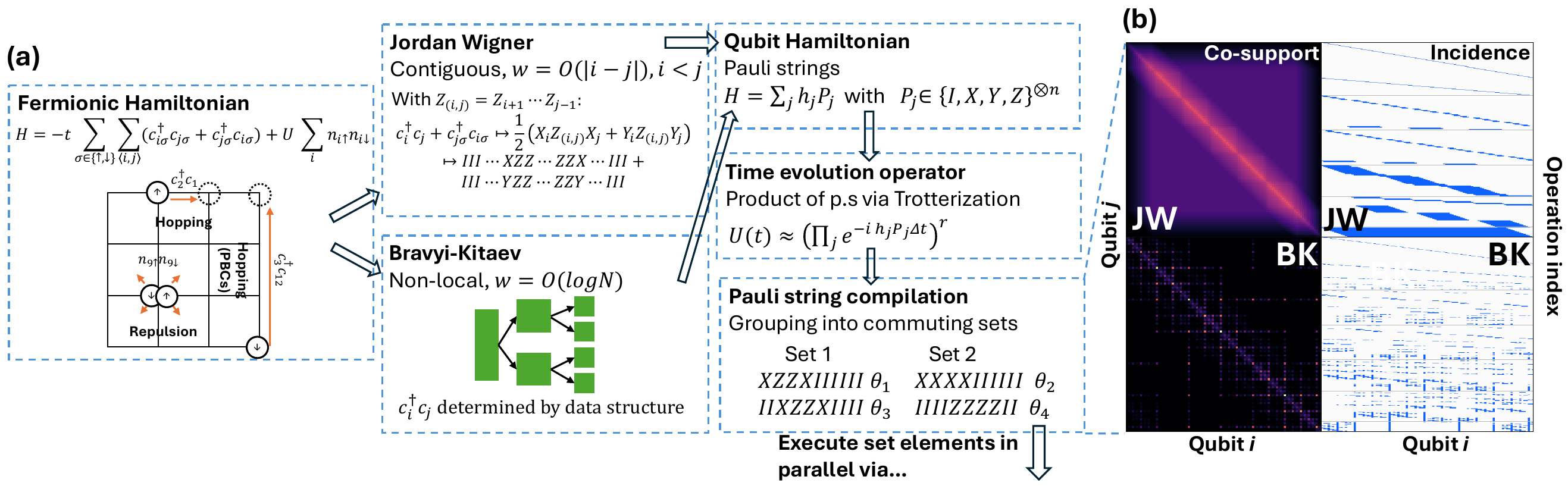}
  \includegraphics[width=0.99\linewidth]{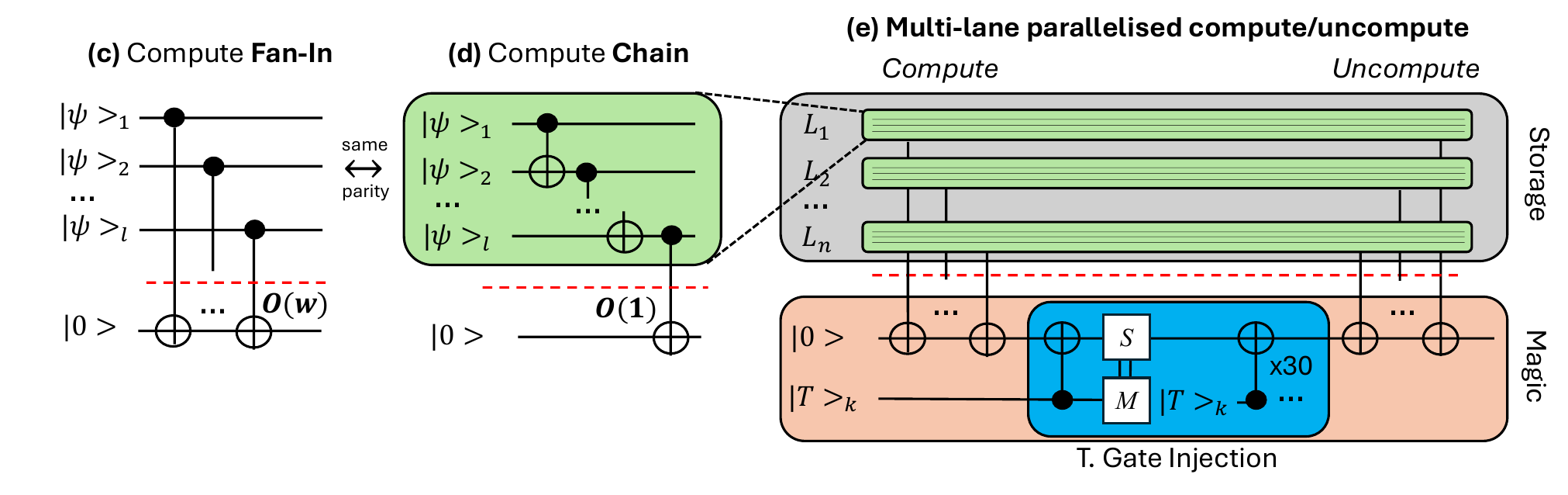}
  \caption{
    \textbf{Simulation workflow, Pauli-string structure and device implementation.}
    \textbf{(a)} Simulation workflow.
    \textbf{(b)} Co-support and incidence matrices for the
    Jordan--Wigner (JW, top) and Bravyi--Kitaev (BK, bottom) 
    transformations on a $7\times7$ Fermi-Hubbard Hamiltonian. The co-support matrices indicate the probability
    that qubits $i$ and $j$ both occur in the support of a Pauli string.
    The incidence matrices indicate which qubits occur in the support of
    each Pauli string.
    \textbf{(c)} A method of computing parity that is suboptimal on
    distributed devices, requiring $O(w)$ non-local CNOTs for a Pauli
    string of weight $w$. 
    \textbf{(d)} A communication-efficient method of computing parity. For each
    lane on which a Pauli string has support, only one non-local CNOT is
    required between the lane and the compute node.
    \textbf{(e)} Logical circuit overview of the implementation of Pauli
    rotations on the distributed device. Single qubit Clifford gates applied during rotation synthesis omitted for clarity. }
  \label{fig:simulation_workflow}
\end{figure*}

Architectural specialization and distribution are promising approaches for scaling fault-tolerant quantum computers. By partitioning computation across modules, a distributed architecture reduces the burden on individual devices and allows memory, computation and communication to be optimized independently. A natural design philosophy~\cite{filippov2026architectingdistributedquantumcomputers, jacinto2026network, Van_Meter_2007, necaise2026distribution, switchQnet, kaur2026impactnetworkconstraintsfaulttolerant} is to equip each module with tightly integrated memory, computation and communication resources, maximizing local capability while minimizing the need for remote operations. However, this organization duplicates expensive resources across the system and can introduce substantial overhead in control, routing, and scheduling. These overheads are particularly significant in early fault-tolerant systems, where distributed communication via Bell pairs may be slow and expensive, making contention for these operations a significant performance bottleneck.

At the same time, fault-tolerant~\cite{shor1996fault,grover1997quantumtelecomputation, gottesman2022opportunities} workloads impose additional architectural requirements. 
Non-Clifford resources~\cite{eastin} must be available where and when they are needed, and the qubit interactions induced by the compiled algorithm must be implemented efficiently. 

Many quantum simulation workloads~\cite{prakash-beverland-resource-estimates, lee2021even} exhibit an asymmetric structure in which parity is accumulated using Clifford operations across a large set of qubits (the support of a Pauli string) but non-Clifford resources are applied only to one of these qubits. This is analogous to a classical workload with high ``arithmetic/operational intensity" \cite{williams2009roofline} which refers to the number of floating-point operations per byte accessed from memory. In the architecture we introduce we can describe quantum simulation tasks as having high magic intensity: the amount of T-gates applied is much higher than the number of accesses from the extensible storage system. The effectiveness of a distributed architecture therefore depends not only on the underlying hardware, but also on how well its organization matches these workload characteristics.

Quantum simulation of electronic systems~\cite{quantumSim, trotter, aspuru2005simulated}, a task where quantum computers are used to simulate the properties of quantum systems including molecules, strongly correlated materials, and lattice models of interacting fermions~\cite{quantumChem, latticeModelRev}, is a particularly useful setting in which to study these trade-offs. The communication structure of these workloads is shaped not only by lattice symmetries and flavor couplings, but also by the chosen fermion-to-qubit mapping. For example, the Jordan--Wigner~\cite{jordan1928paulische} mapping produces geometrically structured but high-weight Pauli strings, whereas Bravyi--Kitaev~\cite{bravyi2002fermionic} yields lighter-weight strings with more non-local and overlapping supports. The suitability of a distributed architecture therefore depends not only on problem size, but also on how the chosen mapping translates fermionic structure into communication, movement, and parallelism requirements. 

Here, we introduce a heterogeneous distributed architecture designed to exploit this structure while maintaining a simple physical organization. The architecture combines a single magic core with one-dimensional chains of cold-storage modules, separating magic-state generation and consumption from logical data storage.  This organization eliminates the need for a complex communication network, reduces the control and calibration burden on storage modules, simplifies routing and scheduling, exposes parallelism across independent storage lanes, and enables the machine to scale by extending simpler storage rather than replicating full execution capability. We note that this architecture can also be realized within a single device. 
We evaluate the architecture on two physically relevant applications, compiled single-band square lattice Fermi--Hubbard~\cite{fh} and sparse Sachdev–Ye–Kitaev (SYK)~\cite{SYK} simulation workloads, which span communication patterns ranging from structured local interactions to highly non-local overlapping Pauli strings. We show that the proposed architecture supports efficient execution of these workloads with experimentally modest module sizes and entanglement-generation rates. We also evaluate the performance of the extensible storage system on random Pauli strings with varying degrees of non-locality. More broadly, our results demonstrate that useful distributed fault-tolerant quantum simulation does not require resource-intensive highly-connected homogeneous modules or complex network fabrics. Instead, efficient execution can be achieved by matching the architecture to the communication structure of the compiled workload.

%% file: arch.tex
\section{Distributed Architectures and quantum simulation}

We consider a local Fermi-Hubbard model~\cite{fh} on a square lattice with periodic boundary conditions  and a non-local sparse SYK model~\cite{SYK}, both ranging in size from 98 to 450 qubits (see Table~\ref{tab:workloads} for all sizes considered). 
These models have broad interest across condensed matter physics, high-energy theory, and quantum information science~\cite{dalzell2025quantum}.
The fermionic Hamiltonians are mapped to qubit operators using either the Jordan--Wigner (JW) or Bravyi--Kitaev (BK) transformations (see Appendix~\ref{sec:ferm_qub_trans} for details)~\cite{jordan1928paulische, bravyi2002fermionic}, and the time evolution operator is implemented using a first-order Trotter-Suzuki~\cite{suzuki1976generalized} decomposition, resulting in a sequence of exponentiated Pauli strings that can be grouped into commuting sets. See Fig.~\ref{fig:simulation_workflow}(a) for a summary of this workflow.
As illustrated in Fig.~\ref{fig:simulation_workflow}(e), executing exponentiated Pauli strings on a quantum computer involves a sequence of two operations: Pauli string compute (uncompute) operations and associated Pauli rotation~\cite{quantumSim}. Compute in this case means using CNOTs to map the parity of qubits supported by an operator onto an ancilla qubit to perform a rotation. Parities can be computed in many ways, some of which are more or less amenable to distributed settings (compare Fig.~\ref{fig:simulation_workflow}(c) and (d)). Uncompute disentangles the qubits from the ancilla after the rotation.

The choice of fermion-to-qubit mapping strongly influences the architectural requirements.
The JW transformation generates nearest-neighbor local qubit operators that are often disjoint and thus readily parallelizable~\cite{maskara2025fast}.
However, the weight of the resulting Pauli strings scales with $O(N)$, where $N$ is the number of fermionic modes. 
By contrast, the BK transformation generates significantly lower weight Pauli strings with $O(\log N)$ scaling, but at the cost of introducing highly non-local Pauli strings, usually with overlapping supports. 

These differences produce substantial architectural consequences even at modest system sizes. For the 98-qubit Fermi–Hubbard instance the average BK Pauli weight (i.e., the non-identity gates in a Pauli string) is approximately 5.7, compared with approximately 11.1 under JW; at 450 qubits, the corresponding averages are approximately 6.6 and 22.8. The benefit of BK therefore grows with problem size, provided that the architecture can efficiently execute its less regular and more spatially distributed supports.
Fig.~\ref{fig:simulation_workflow}(b) illustrates these differences: the co-support matrix of JW is centered around the diagonal, indicating that neighboring strings feature qubits close to each other. 
At the same time, the matrix is densely populated due to the presence of long Pauli strings that involve almost all qubits. For the case of the Fermi-Hubbard model, the longest strings come from the periodic boundary conditions (PBCs). This is because fermionic nearest-neighbor at the lattice boundaries cross the entire lattice (they are wrapped around) and in JW weight is proportional to the distance between sites in the lattice (assuming standard ordering). 

In contrast, the BK cosupport matrix is substantially sparser but with significant cosupport between distant qubits. This is because BK represents these long-range hopping terms with a tree (typically Fenwick \cite{fenwick}) encoding, which saves weight but produces non-locality.
The corresponding incidence matrices further highlight the difference in regularity. 
JW produces highly regular lines of support, whereas BK generates sparse and irregular support.

Historically, the JW transformation has been preferred for fault-tolerant simulation because of the difficulty of accessing and interacting highly non-local supports~\cite{maskara2025fast, necaise2026distribution}. 
However, the long Pauli strings generated by the JW transformation can be expensive to execute, incurring large movement costs for larger systems and more complex fermionic models, especially on distributed architectures. 

Conversely, although BK substantially reduces the Pauli-string weights, the high communication costs (evaluated in Sec.~\ref{sec:storage}) associated with non-local operators with supports spread across module(s) can offset these benefits, especially in distributed architectures where different and typically more expensive modalities (Bell pairs, shuttling, long movements, swaps, etc) are needed to interact with distant qubits. 
This tradeoff motivates the development of architectures that can efficiently support non-local Pauli strings without incurring prohibitive communication overheads.

\begin{figure}[t!]
  \centering
  \includegraphics[width=0.99\linewidth]{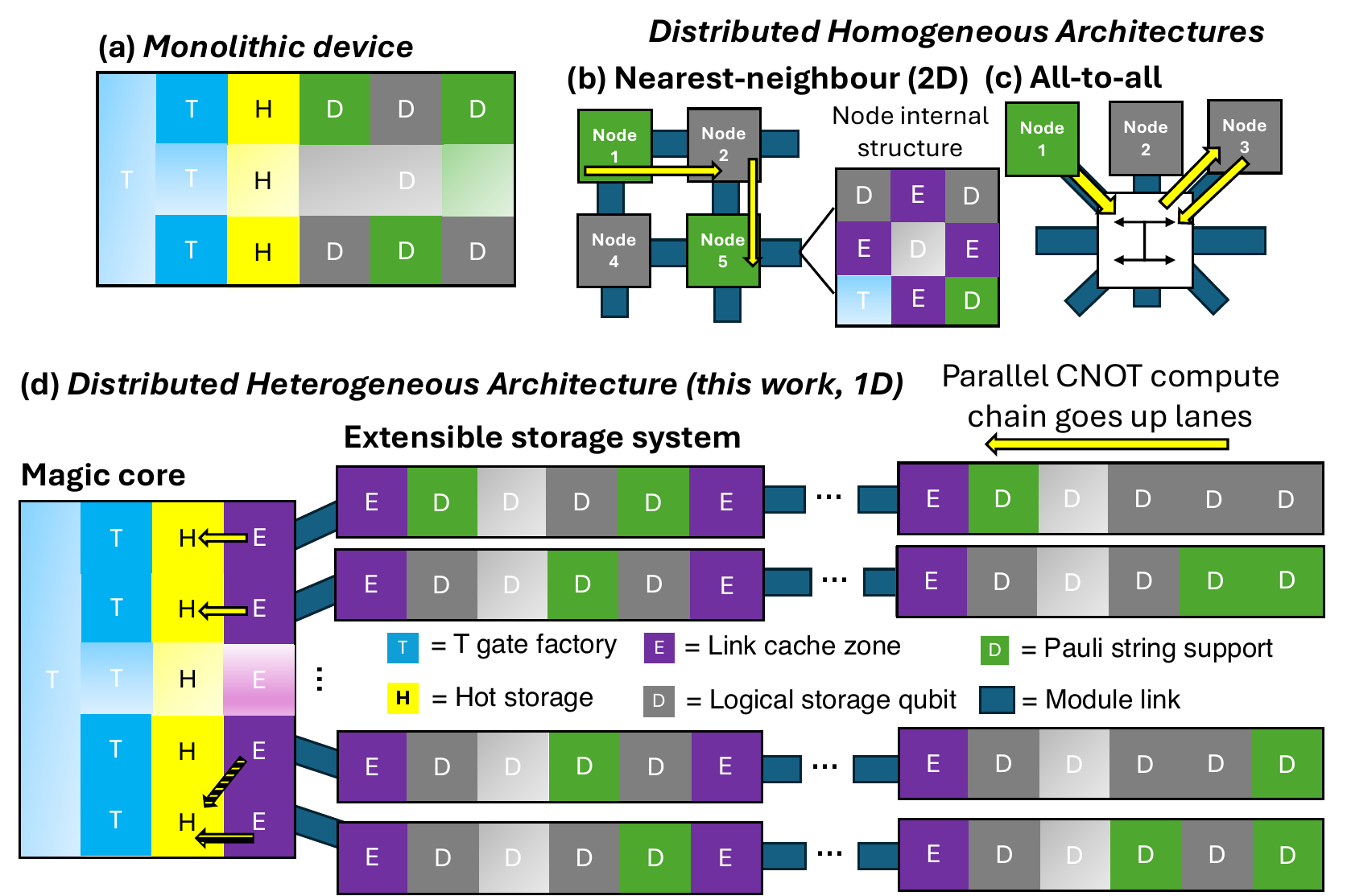}
\caption{\textbf{Quantum architectures.} Gradient coloring represents design space for sizing. \textbf{(a)} Monolithic device with column(s) of T factories (T), hot-storage zones (H), connected via ebits (E) to data (D) / cold storage qubits. \textbf{(b-c)} Distributed homogeneous architectures. Yellow lines indicate node-to-node interactions required to implement distributed Pauli strings. Example internal homogeneous node structure shown. \textbf{(d)} Distributed heterogeneous architecture. Yellow and cross-hatched yellow arrows at the Magic core indicate inter- and intra-string parallelism respectively. Note that while nodes are shown as squares or 1D rectangles, this is just illustrative. We model nodes as rectangles with a standard aspect ratio. }
\label{fig:arch}
\end{figure}

To address this challenge, we propose a heterogeneous lane architecture shown in Fig.~\ref{fig:arch}(d). The architecture consists of a centralized magic core connected to multiple cold-storage lanes.
The magic core hosts the resources required for active execution, including T-state factories (T), hot-storage zones (H) for frequently accessed logical qubits, and a communication zone for caching Bell pairs (E) and for connection to cold-storage lanes consisting of modules which hold data (logical) qubits (D). The cold-storage lanes contain the majority of logical data and are extended at low relative cost as required to accommodate larger problem instances, without increasing network/bus complexity. This organization separates active computation from long-lived logical storage and allows each component of the system to be optimized for its specific role. 

A key advantage of this design is that communication contention is largely eliminated: each lane in the heterogeneous architecture is dedicated to the Pauli-string task currently being executed.
Consequently, the communication cost is largely independent of the spatial distribution of the Pauli-string support. For comparison, we also evaluate a homogeneous distributed architecture (Fig.~\ref{fig:arch}(b), the ``2D design''), which requires substantially greater communication resources. While all-to-all or highly connected approaches (Fig.~\ref{fig:arch}(c)) have been proposed~\cite{doi:10.1126/science.1231298}, these have high experimental costs, with approaches such as including a switch in the optical generation of Bell pairs generating infidelities~\cite{stackSwitch, ciscoSwitch}. Maintaining Bell pair caches in an all-to-all architecture would mean that each device must have $O(N_{nodes})$ Bell pair caches, which would take up a significant amount of space.

Another key consideration in our architectural model is that magic state generation is substantially more resource-intensive than logical memory.
Devices that generate magic states have significantly greater complexity and usage requirements, as they must continuously produce magic states during computation. This means that a substantial fraction of the device must be dedicated to T-state factory operation, limiting the resources available for other functions. In current neutral-atom platforms, for example, the number of Acousto-Optical Deflectors (AODs) is limited, restricting the degree of parallelism that can be achieved while requiring a significant fraction of the available control resources to be devoted to T-state production.
By separating magic-state generation from logical storage, the proposed architecture avoids replicating these expensive resources across every module, enabling the system to scale by adding low-cost storage modules rather than fully featured compute nodes. 

In what follows we will assume logical information to be encoded using surface-code patches~\cite{litinski2019game}, with logical operations implemented transversally~\cite{shor1996fault} and via magic state injection. This implementation is well suited to reconfigurable neutral-atom arrays~\cite{bluvstein2024logical} and trapped-ion devices~\cite{ryan2024high}, where logical patches can be moved and coupled efficiently. However, the extensive encoding overhead from surface codes lead to different challenges in trapping and routing logical qubits across long distances; in this case, having smaller nodes and distributed logic comes again as an advantage.

For the quantitative evaluation in this work, we consider a neutral-atom implementation and adopt standard experimental device parameters from the literature~\cite{zhou2025resource}, shown in Tab.~\ref{tab:params}. 
These values are chosen to reflect current experimental capabilities rather than to optimize performance for the proposed architecture.

%% file: related.tex
\section{Related work}

In fault-tolerant architecture design, the idea of specialized (heterogeneous) sections of a device is not new. These sections may use different QEC codes for different purposes (cold storage or using magic)~\cite{gidney2025factor, hetEC} or use identical QEC codes but with dedicated sections for minimizing contention and routing costs~\cite{lee2021even}. The idea of distributing devices and designing around local and non-local connections is also not new, or even recent~\cite{grover1997quantumtelecomputation, Cirac_1999, Van_Meter_2007}. Further, many fault-tolerant architectures involve some form of parallel processing with multiple sources of magic~\cite{zhou2025resource, webster2026pinnaclearchitecturereducingcost}.

However, prior work generally considers homogeneous distribution, where the same device is replicated multiple times and interconnected with a simple~\cite{filippov2026architectingdistributedquantumcomputers, jacinto2026network, Van_Meter_2007, necaise2026distribution} or complex~\cite{switchQnet, kaur2026impactnetworkconstraintsfaulttolerant} network. The distinction of our work is that our architecture is heterogeneous but also distributed and extensible, and we show that our design performs better than a simple homogeneous distributed device for matched T factory count. The fact that our device is distributed means that it can be arbitrarily extended to match the size of the workload executed. The heterogeneity and simple network design ensures the extensibility is low-cost: less-complex storage nodes are simply added to the chain. A recent work has started exploring heterogeneous distributed architectures~\cite{qctrl}. But their proposal is less specialized than our architecture and relies on a centralized quantum bus to provide all-to-all logical connectivity.

Prior work has also considered using reconfigurable architectures and dynamic encoding changes to perform fermionic simulation efficiently~\cite{maskara2025fast,aigner2026fermionlatticessimulatedsamesize,li2026asymptoticallyoptimaldepthfermionic}. These approaches are very promising for monolithic devices and discuss compatibility with fault-tolerant execution, but the dynamic encoding changes come with costs. The low-depth approach in~\cite{maskara2025fast} uses $O(N)$ ancilla qubits, the strongest results in~\cite{aigner2026fermionlatticessimulatedsamesize} apply to geometrically local or bounded-range fermionic systems, and the arbitrary fermionic permutations in~\cite{li2026asymptoticallyoptimaldepthfermionic} require $O(\sqrt{N})$ depth and $O(N\sqrt{N})$ gates. However, none of these explicitly model a complete and distributed fault-tolerant architecture with magic-state production, modular communication, and contention. Furthermore, the required global reconfiguration or routing is natural within a monolithic device, but would require inter-module communication if the data were distributed across separate modules, a cost that these works do not consider. Our work instead takes the compiled Pauli strings as given and considers how to execute their local and non-local supports on an extensible distributed architecture. Explicit architectural studies of dynamic encodings on distributed devices are a promising direction for future work.

\begin{table}[t!]
\centering
\begin{tabular}{ll}
\hline
\textbf{Parameter} & \textbf{Value} \\
\hline
T-gate generation speed per factory & $10\,\mathrm{ms}$ \\
T-gates per rotation & $30$ \\
Bell pair generation speed & $1\,\mathrm{ms}$  \\
Bell pair parallelism & $O(d)$ \\
Bell pair cache size & $n=d^2$ \\
Neutral atom movement acceleration & $5500\,\mathrm{m\,s^{-2}}$ \\
\hline
\end{tabular}
\caption{\textbf{Parameters used in the architecture model.}}
\label{tab:params}
\end{table}

%% file: eval.tex
\section{Evaluation Methodology}

\subsection{Trotterization}
After the fermion-to-qubit transformation, each Hamiltonian is expressed as a sum of
Pauli operators,
{
\setlength{\abovedisplayskip}{4pt}
\setlength{\belowdisplayskip}{4pt}
\begin{equation}
    H = \sum_{j=1}^{M} h_j P_j,
\end{equation}
}
where $h_j$ are real coefficients and $P_j$ are tensor products of Pauli
operators, also referred to as Pauli-strings. $M$ is the total number of Pauli strings per Trotter step.
The time-evolution operator for evolution time $t$ is $U(t) = e^{-iHt}.$

We approximate the evolution using a first-order Trotter-Suzuki decomposition~\cite{suzuki1976generalized}. 
Dividing the total evolution time into $r=10$ equal steps of duration $\Delta t=t/r$ gives
\begin{equation}\label{eq:propagator}
    U(t)
    =
    e^{-iHt}
    \approx
    \left[
    \prod_{j=1}^{M}
    e^{-i h_j P_j \Delta t}
    \right]^r
    + O( t^2/r).
\end{equation}

Every exponential term $e^{-i h_j P_j \Delta t}$ corresponds to a Pauli rotation and is implemented using the standard compute-rotate-uncompute procedure. Pauli strings produced after encoding fermions into qubits are a mix of Pauli operators on different qubits, so single-qubit Clifford gates are first used to put each qubit in the support of $P_j$ into the $Z$ basis and then the parity of these qubits is accumulated onto an ancilla qubit using a sequence of CNOT gates.
A single-qubit rotation with angle $\theta_j = 2 h_j \Delta t$
is then applied to the ancilla, after which the parity computation is reversed (uncompute). 

We assume that every single-qubit rotation needs 30 T-gates, regardless of architecture. 
Based on recent work~\cite{synth}, this is sufficient for a rotation error of $10^{-6}$ per rotation and any angle, where 40 T-gates suffice for a rotation error of $10^{-8}$.  
We ignore the cost of the single-qubit Clifford gates as they are negligible compared to the other costs in this paper.

All the results and reported execution times correspond to a evolution with a single Trotter step ($r=1$ in Eq.~\ref{eq:propagator}).
However, for calculating the distance estimates for the surface code, we consider the total evolution consisting of $r=10$ Trotter steps (see Appendix~\ref{sec:cdestimates} for details). 

\subsection{Architectures}
We compare our heterogeneous lane architecture with two baseline architectures: a monolithic architecture (Fig.~\ref{fig:arch}(a)) and a homogeneous distributed architecture (Fig.~\ref{fig:arch}(b)).
 As mentioned earlier, we assume the use of neutral atom device(s) and use the parameters shown in Table~\ref{tab:params}. 
 For a movement of a code block with code distance $d$ along dimensionless length $\ell$, we assign a physical distance $L_{\mathrm{phys}} = \ell d \left(12\,\mu\mathrm{m}\right)$ and movement time
\begin{equation}
    \tau_{\mathrm{move}} = 2\sqrt{\frac{L_{\mathrm{phys}}}{a}},
\end{equation}
where, $a=5500\,\mathrm{m\,s^{-2}}$ is the acceleration and $12\,\mu\mathrm{m}$ comes from the atom spacing~\cite{zhou2025resource}.
The same movement model is used for both the heterogeneous architecture and the homogeneous 2D baseline. 

\subsubsection{Heterogeneous lane architecture (1D)}

Logical data qubits are divided into lanes, for example an architecture with 4 lanes and 100 data qubits would have qubit $q_0$ to $q_{24}$ on the zeroth lane, $q_{25}$ to $q_{49}$ on the next lane and so on. Each storage lane is then partitioned into cold-storage modules. 
 Within each lane, qubits fill consecutive storage modules up to the 20-qubit capacity; if the number assigned to a lane is not divisible by 20, its final module contains only the remaining qubits. Unless otherwise stated, in this work this architecture uses six storage lanes, six compute rows, and five T-factory columns ($6\times5=30$ T factories in total).

Internal CNOTs between two qubits in the same module are implemented by moving the control qubit to the target and back, giving a total travel distance equal to twice the Manhattan separation between the qubits. 
Logical operations involving qubits located in different storage modules require non-local CNOT operations, carried out using cached Bell pairs. 
If the two supports on a lane are separated by multiple modules, then entanglement swapping is used to create a Bell pair link via the intermediate modules. 

For a Pauli string, all local CNOTs required to accumulate parity across its supports are performed simultaneously across the storage modules. Then non-local CNOTs are used to accumulate the parities across a lane and feed it into the Magic core.

Pauli rotations are implemented using a compute-rotate-uncompute procedure. 
For a Pauli string $P$, the support of $P$ is first partitioned according to storage lane.
Within each occupied lane, parity is accumulated locally and transferred to the magic core using a non-local logical CNOT operation.
The magic core combines the parities arriving from all occupied lanes and performs the required Pauli rotation. 
The compute step is then applied again but in reverse, which forms the uncompute step.
Consequently, both local and non-local Pauli strings are executed using the same architectural primitive: lane-local parity accumulation followed by centralized rotation.

\subsubsection{Monolithic architecture}
The monolithic baseline represents an idealized monolithic quantum computer. It is the heterogeneous lane (1D) architecture but without communication costs. This is implemented in the simulator by giving the 1D architecture instantaneous Bell pairs.
Consequently, the monolithic architecture provides a lower-bound for execution time, isolating the effects of available parity-computation resources and T-state production from communication constraints.

\subsubsection{Homogeneous distributed architecture (2D)}

The homogeneous distributed device consists of a nearest-neighbor two-dimensional grid of identical modules. Each module contains up to 20 logical data qubits and 2-4 Bell pair/link zones. The number of local \(T\)-state factories is varied in this work. For a
problem containing \(Q\) logical data qubits, the minimum number of
modules required by storage capacity is $m = \left\lceil \frac{Q}{20} \right\rceil$.
We instantiate a complete near-square rectangular grid with
\[
    X = \left\lceil \sqrt{m} \right\rceil,
    \qquad
    Y = \left\lceil \frac{m}{X} \right\rceil,
    \qquad
    N_{\mathrm{grid}} = XY .
\]
Thus, \(N_{\mathrm{grid}}\) may be slightly larger than \(m\). The \(Q\) data qubits are distributed
across all \(N_{\mathrm{grid}}\) modules in snake order.

For the 2D device we use a snake schedule, qubits on modules that are the support of Pauli string are first combined locally in parallel. Non-local CNOTs across the supporting modules are then used to finish the compute step. We explored other routing methods but they increased the number of hops between modules, increasing wall-clock time due to communication cost.

The T-state operation is performed at the module of the ancilla qubit and the compute process is then repeated to uncompute the parity. All support and transit modules used by the network/bus remain reserved throughout this process.

\begin{table}[!t]
\centering
\begingroup
\small
\setlength{\tabcolsep}{4.5pt}
\renewcommand{\arraystretch}{1.0}
\begin{tabular}{@{}lrrrrrrrr@{}}
\hline
Model & $L$ & $n_q$ & Groups & Strings & $W_{\rm BK}$ & $W_{\rm JW}$ & $d_{\rm BK}$ & $d_{\rm JW}$ \\
\hline
FH    & 7  & 98  & 6  & 539  & 3080   & 5964   & 12 & 12 \\
FH    & 10 & 200 & 5  & 1100 & 6630   & 17040  & 12 & 13 \\
FH    & 12 & 288 & 5  & 1584 & 9816   & 29184  & 13 & 13 \\
FH    & 15 & 450 & 6  & 2475 & 16254  & 56460  & 13 & 14 \\
SYK & -  & 98  & 7  & 129  & 2028   & 5394   & 10 & 11 \\
SYK & - & 200 & 9  & 294  & 5713   & 24366  & 11 & 13 \\
SYK & - & 288 & 11 & 419  & 8873   & 50335  & 12 & 13 \\
SYK & - & 450 & 9  & 651  & 15589  & 117871 & 12 & 14 \\
\hline
\end{tabular}
\endgroup
\caption{\textbf{Quantum simulation workloads.} We report the system size L (for FH only), the number of logical qubits $n_q$, the number of commuting groups, the total number of Pauli strings, the sum of Pauli weights across all strings $W_{BK}, W_{JW}$, and the estimated code distances $d_{BK}, d_{JW}$ used in the wall clock time simulations for the different mappings.
Here, $W$ is the total Pauli weight, i.e., the sum of support sizes over all non-identity Pauli terms.
}
\label{tab:workloads}
\end{table}

\begin{table}[t!]
\centering
\begingroup
\small
\setlength{\tabcolsep}{2.8pt}
\renewcommand{\arraystretch}{1.0}
\resizebox{\columnwidth}{!}{%
\begin{tabular}{@{}lrrrrrrr@{}}
\hline
Arch. & N & Q/Module & Data & Hot & Bell & $T$ & Total \\
\hline
Mono. 20L/5T & 1 & 68,970 & 54,450 & 2,420 & 0 & 12,100 & 68,970 \\
1D 6L5T & 1+24 & 5,082/2,511 & 54,450 & 726 & 6,534 & 3,630 & 65,340 \\
2D 1T & 25 & 2,807 & 54,450 & 3,025 & 9,680 & 3,025 & 70,180 \\
2D 5T & 25 & 3,291 & 54,450 & 3,025 & 9,680 & 15,125 & 82,280 \\
\hline
\end{tabular}%
}
\endgroup
\caption{ \textbf{Physical-qubit footprint for 450-logical-qubit workloads for each architecture.} We fix $d=11$, differences in encoding choice can lead to slightly different distances.}
\label{tab:qubits}
\end{table}

\subsection{Non-local operations and Bell pair generation}

Our simulations are specific to the case where communication between modules (non-local operations) is via Bell pairs. But the results and analysis are broadly relevant to architectures where communication between subsystems is slower than within, and especially where some form of caching can be used. 

In the heterogeneous architecture, each communication link maintains independent Bell pair pools for the two possible communication directions. In the homogeneous architecture, 2-4 pools are maintained depending on whether a module is on an edge or in the bulk.

For a code-distance $d$ simulation, each external logical operation consumes $n=d^{2}$ Bell pairs from the corresponding pairs of pools. This is because a non-local CNOT via transversal operations requires one Bell pair per physical CNOT, and a transversal operation uses $n$ physical CNOTs between the data qubits of the two codeblocks~\cite{Stack_2026}. Importantly, provided the Bell-pair fidelity is within an order of magnitude of the local physical error rate, the logical error rate of a non-local operation is exponentially suppressed with respect to code distance~\cite{Stack_2026}, making error suppression possible. Detailed consideration of fidelities and decoherence are outside of the scope of this work.

We assume that Bell pairs are generated continuously at a rate $\tau_{B} =d$ Bell pairs per $\mathrm{ms}$ per directed pool. The parallel generation of $O(d)$ atom-photon pairs has been demonstrated experimentally in~\cite{covey}. We take parallel generation of Bell pairs as an architectural assumption. If parallel generation is infeasible, the results of our paper can be adapted to serial ebit generation by multiplying the generation time by $O(d)$.  We choose $O(d)$ Bell pair generation times of 1ms and 10ms. The 1ms time is roughly equal to syndrome extraction cycle time on a neutral atom device~\cite{zhou2025resource} and is experimentally feasible~\cite{oxfordNetwork} for non-local implementations. The 10 ms time represents a pessimistic scenario in which the Bell state generation time is far slower than the QEC cycle time.

\subsection{Pauli rotation and T-state generation}

As described above, each Pauli rotation is assigned a fixed non-Clifford cost of 30 T states~\cite{Kliuchnikov_2023}. T states are supplied by dedicated T-state factories located within each module of the homogeneous architecture, and within the magic core of the heterogeneous architecture. 
A single T-factory produces one T state every 10 ms. Consequently, a Pauli rotation requiring 30 T states has a nominal T-stage time $\tau_T = \frac{300}{n_T} \mathrm{ms}$,
where $n_T$ is the number of available T-factory columns.
 In general, implementing these rotations requires a string of $T$ and Clifford gates. Again, we ignore the cost of the Clifford gates as this is specific to the method of rotation synthesis used and is relatively negligible compared to the other costs in this paper. 
In the homogeneous architecture, each module has the same factory count. In the heterogeneous architecture, unless otherwise stated, we use $n_T=5$ columns of T factories. This means each storage lane has five T factories dedicated to it. This corresponds to a nominal T-stage duration of 60 ms for (up to) the number of Pauli rotations equal to the number of lanes. A Pauli-string task reserves a compute row for its entire lifecycle, including parity accumulation, T-state consumption, and parity uncomputation.

\subsection{Grouping}
Given the Hamiltonian expressed as a sum of Pauli operators, the time evolution operator is constructed by partitioning its Pauli terms into commuting groups~\cite{anand2025hamiltonian,anand2025leveraging,van2020circuit}, with the groups executed sequentially.

The Pauli operators are partitioned into commuting groups using Qiskit's~\cite{qiskit2024}
\begin{center}
\texttt{group\_commuting(qubit\_wise=False)}
\end{center} routine. 
Operators belonging to the same group commute and therefore can be executed in any order, while operators with disjoint supports can additionally be executed simultaneously. Once every operator in a group has been executed, the next group can start.

Within a group, the cycle-level scheduler greedily starts all Pauli strings whose qubit supports and lane sets are disjoint from currently active strings. 
Thus the grouping determines a legal ordering of mutually commuting rotations, while the simulator still enforces architectural conflicts at the level of qubits, lanes, and storage modules.

\subsection{Simulation workloads}
We consider systems of sizes ranging from 98 to 450 logical qubits for both the Fermi-Hubbard and sparse SYK models.
Table~\ref{tab:workloads} summarizes the properties of the workloads considered in this work.

\subsection{Scheduling model}

The simulator advances all active objects in discrete cycles. 
At the beginning of each commuting group, all Pauli strings belonging to that group are placed in a scheduling queue.
The scheduler greedily scans the queue and launches a Pauli-string task whenever its logical-qubit support and storage-lane set are disjoint from those of all currently active tasks.
Strings that conflict with active tasks remain queued until the required resources become available.

Once a task has been launched, each occupied storage lane executes its parity-accumulation procedure independently.
Consequently, the occupied lanes of a single Pauli string can operate in parallel, while multiple disjoint Pauli strings within the same commuting group are also executed concurrently.

A task completes only after all occupied lanes have finished uncomputation, at which point its logical qubits, storage lanes, communication resources, and assigned row of T factories are released.

%% file: perform.tex
\begin{figure}[t!]
  \centering
  \includegraphics[width=0.99\linewidth]{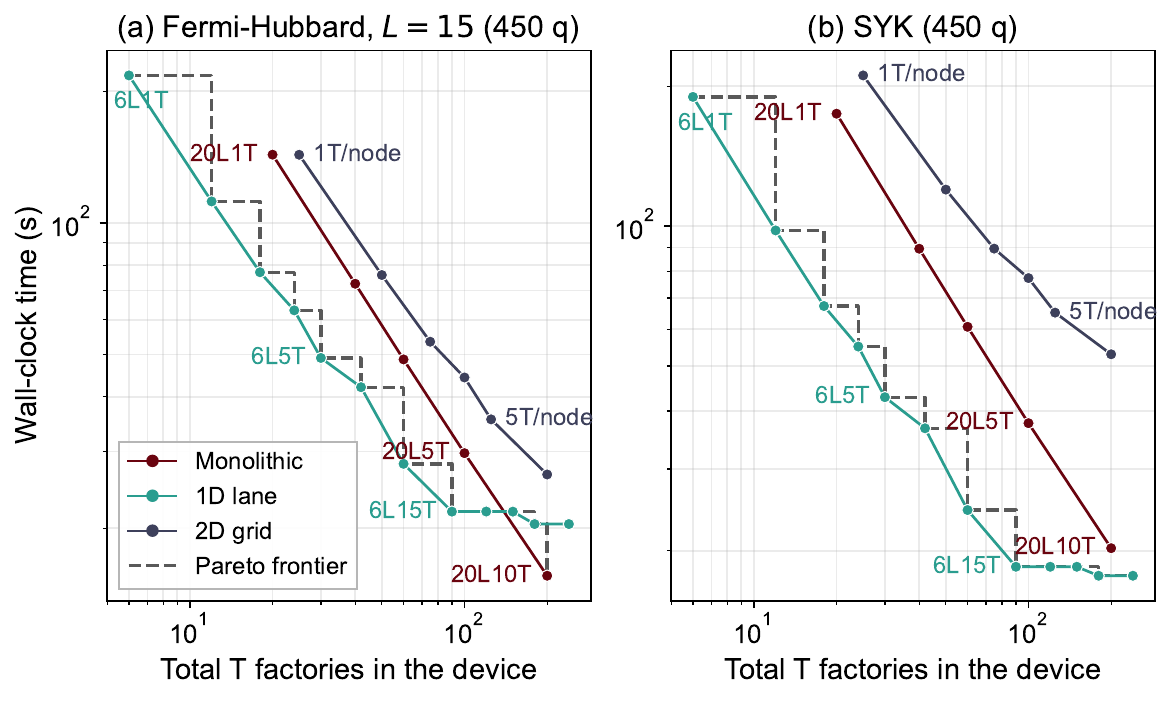}
  \caption{\textbf{Wall-clock time against total T factory count Pareto plot.} \textbf{(a-b)} 450-qubit simulations at 1ms $O(d)$ Bell state generation rates. The data shown in these plots correspond to the Bravyi--Kitaev (BK) transformation. }
  \label{fig:pareto}
\end{figure}

\begin{figure}[t!]
  \centering
    \includegraphics[width=0.99\linewidth]{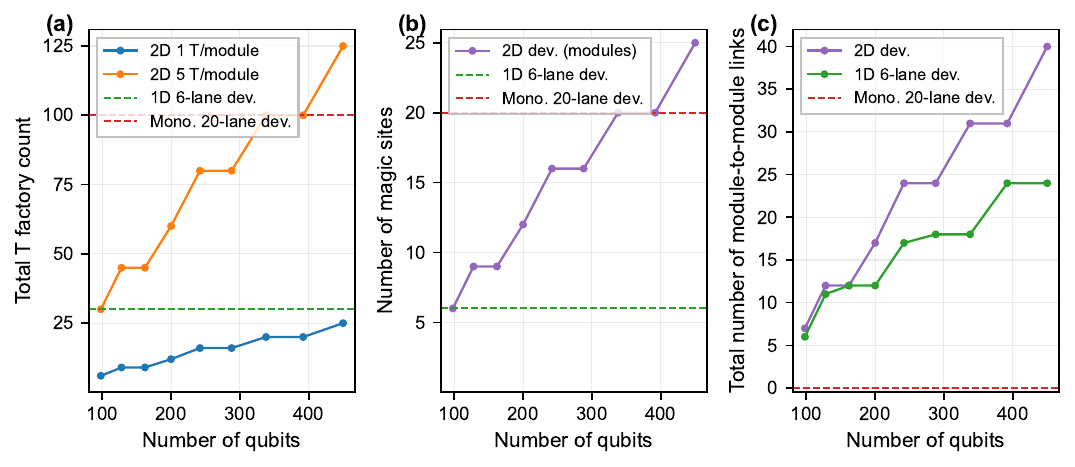}
  \caption{\textbf{Number of T factories, magic sites and distributed connections.} \textbf{(a)} Total number of T factories across different distributed quantum computing architectures as problem size increases.  \textbf{(b)} Total number of magic sites for each architecture. For the 1D and monolithic architectures, this remains fixed and is equal to the number of storage lanes. For the 2D architecture, it is equal to the number of modules and therefore increases with problem size. \textbf{(c)} Total number of distributed connections for different architectures against qubit count. }
  \label{fig:computeSites}
\end{figure}

\begin{figure}[htpb!]
  \centering
  \includegraphics[width=0.99\linewidth]{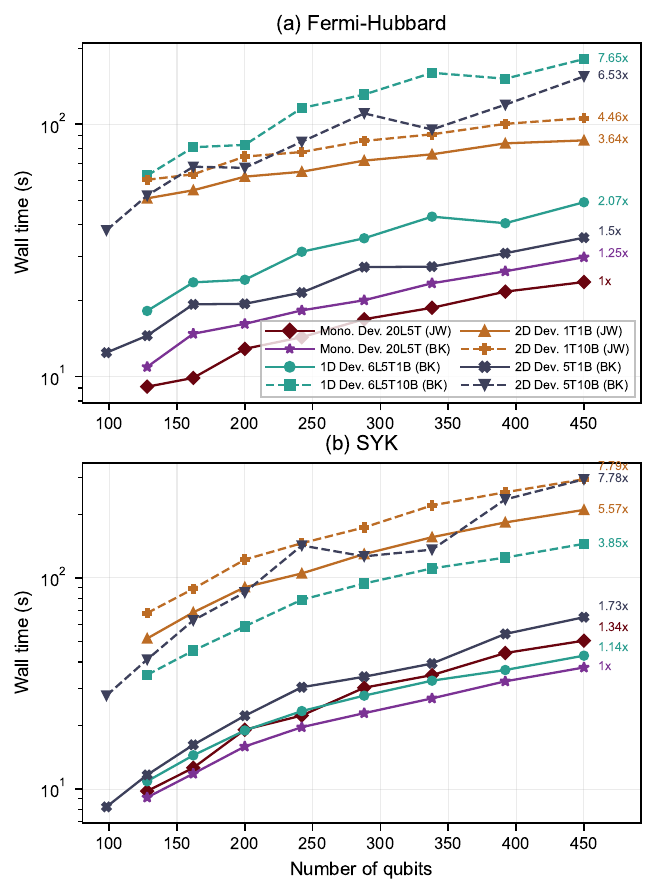}
  \caption{\textbf{Performance of the architecture and its scaling against problem size.} \textbf{(a-b)} The performance of the device on an $L\times L$ grid Fermi-Hubbard simulation task with $N=2L^2$ and a Sparse SYK simulation task with $N$ complex fermionic modes. $aL$ refers to $a$ lanes, $bT$ refers to $b$ $T$-factory columns, and $cB$ refers to $O(d)$ Bell pair generation time of $c$ ms. End point ratios are against the fastest device.}
  \label{fig:overall}
\end{figure}

\section{System performance}

\begin{itemize}

\item \textbf{Observation 1: With substantially fewer magic sites and T-state factories, the heterogeneous (1D) architecture approaches monolithic and 2D device performance.} 
\item \textbf{Observation 2: For matched T factory count and under the same encoding, the 1D device is almost always faster than both the monolithic and 2D architectures.}

\end{itemize}

\begin{figure*}[htpb!]
  \centering
  \includegraphics[width=0.99\linewidth]{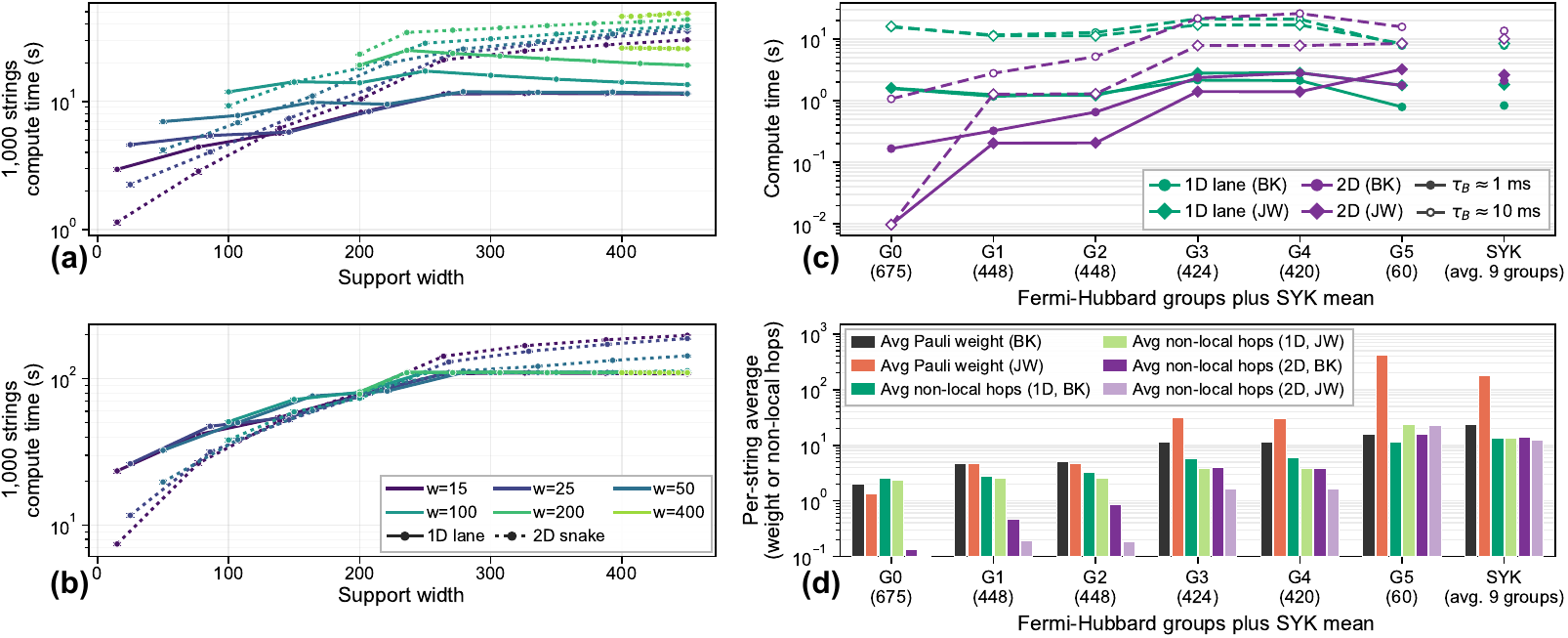}
  \caption{\textbf{Compute time for random and work-load specific Pauli strings on different architectures.}  \textbf{(a-b)} Random simulations. We fix $d=11$ here. Each graph shows the total time taken to execute 1000 strings of weight $w$ where for each Pauli string the supports are uniformly randomly distributed across a set of qubits [$c-W/2, c + W/2 $] of width $W$, where $c$ is randomly chosen for each string. Error bars from re-running the experiment are smaller than the data points. \textbf{(a)} 1 ms $O(d)$ Bell pair generation speed. \textbf{(b)} $10\times$ slower $O(d)$ Bell pair generation time (10 ms). \textbf{(c)} Time to compute the different commuting groups of BK strings in a 450-qubit Fermi-Hubbard simulation for 1 ms and 10 ms $O(d)$ Bell pair generation times. We also report the results for a 450-qubit SYK simulation. This is averaged across the groups because the results were very similar for each group. \textbf{(d)} Average Pauli weight for BK and JW per Fermi-Hubbard commuting group (or SYK across all groups) and the average number of non-local hops in the different architectures and compilation methods.  }
  \label{fig:storage}
\end{figure*}

Figure~\ref{fig:pareto} shows the performance--resource Pareto frontier for the architectures considered in this work as the total $T$ factory count is varied. The six-lane heterogeneous (1D) architecture lies on the Pareto frontier across most resource budgets, only being overtaken by the 20 lane equivalent monolithic architecture once its 6 execution lanes become saturated. Importantly, the homogeneous distributed (2D) architecture never outperforms the 1D design for the same number of $T$ factories. The performance gap is very significant for the SYK workload. While the 2D architecture for matched number of qubits, $n_q$, has more magic sites for parallel processing (see Fig~\ref{fig:computeSites}), the strings in this workload are difficult to execute in parallel due to their weight and disjointness (see Sec.\ref{sec:storage} and Sec.\ref{sec:para} for more details). 

For the rest of this paper, we will focus on comparisons to the six lanes with five T-factories (6L5T) 1D architecture. This is because whilst its magic core has a larger physical qubit count (Tab.~\ref{tab:qubits}) of $\approx 5100$ compared to a 2D architecture module ($\approx 3300$) with five T-factories, the storage modules in the 1D device have on average $\approx2500$ physical qubits.

Figure~\ref{fig:overall}(a) shows the performance of different architectures for Fermi-Hubbard simulations across a range of lattice sizes. 
Despite having only six parallel lanes that have five $T$ factories each, the heterogeneous architecture is only about $2.1\times$ slower than the much larger monolithic baseline. This corresponds to executing 1 Trotter step in $49.1$s.
If execution scaled solely with the number of available lanes, one would expect a slowdown of approximately $3.3\times$.
The smaller observed slowdown indicates that this simulation workload does not contain sufficient parallelism to fully utilize the larger monolithic device, even using the JW encoding.

With five $T$ factories per node, the homogeneous architecture executes a Trotter step in $\approx 35.5$s, $1.5\times$ slower than the monolithic device and $1.4\times$ faster than the 6L5T device.  
This is faster than the heterogeneous architecture. However, this performance comes from having approximately 125 $T$ factories in total, which is more than the 100 $T$ factory monolithic device, and $4\times$ more than the 6L5T heterogeneous architecture, which has only 30 $T$ factories (Fig.~\ref{fig:computeSites}(a)).
When restricted to one $T$ factory per node, corresponding to approximately 25 factories in total and therefore a resource budget comparable to that of the heterogeneous architecture (Fig.~\ref{fig:computeSites}(a)), the homogeneous architecture is $\approx2\times$ slower than the 1D device. Note that for the 2D 1T FH case, JW is used as it enables better use of the more limited magic compared to the 5T case, even if it induces a higher movement cost.
In both cases, the homogeneous design also requires substantially greater communication connectivity and routing resources. 

To evaluate sensitivity to communication performance, we reduce the generation speed of $O(d)$ Bell pairs from 1 ms to 10 ms.
The execution time of the heterogeneous architecture nearly quadruples. However, the degradation in performance is substantially less than the $10\times$ drop that would be expected.

Figure~\ref{fig:overall}(b) presents equivalent results for sparse SYK simulations. Here, as noted before, the 1D device performs consistently better, with the gap between it and the 2D device growing with qubit count. The gap widens for slower Bell state generation rate.

%% file: storage.tex
\section{Extensible storage performance}
\label{sec:storage}

\textbf{Observation 3: The 1D lane architecture becomes more competitive as Pauli-string support becomes more spatially non-local.}

\textbf{Observation 4: Slower Bell pair generation reduces sensitivity to Pauli weight because communication (rather than local parity accumulation) becomes the dominant cost.}

Fig.~\ref{fig:storage} isolates the performance of the extensible storage system by excluding the Pauli rotation step and measuring only the time taken to compute Pauli strings. We find that only at the lowest weight and support width (high locality) does the 2D homogeneous architecture perform better than the 1D heterogeneous architecture. In this case, most Pauli strings are supported within a single module. We see that as the width of the distribution of supports goes up (increasing non-locality), the relative performance of the 1D system increases, becoming quickly better than the 2D device. 

The 1D device's relative performance is best for the most distributed strings. This is because each of these strings are distributed across most or all of the 6 lanes of the device. This means they will be computed using all 6 lanes. Whereas for medium-distributed high-weight strings, it is difficult to continuously form commuting sets with disjoint supports that use all 6 lanes. This is why in Fig.~\ref{fig:storage}(a) the runtime increases from low to medium support widths and then decreases again.

In Fig.~\ref{fig:storage}(b) we see that decreasing the generation rate of Bell pairs by $10\times$ results in clustering behavior between Pauli strings of significantly different weights. 

This is because the cost of internal movement becomes much lower so the difference between high weight and low weight strings that are both widely distributed (and therefore have similar external communication requirements) becomes minimal. 

We also evaluate compute times, average Pauli weights and number of non-local hops under different encodings for the different commuting groups in Fermi-Hubbard simulation (and averaged across the entire simulation for SYK). First, we see how the behaviors identified in the random benchmark above continue in the real simulation: The 1D device is competitive for the more distributed commuting groups. That being said, the disjointness (and therefore parallelism) granted by the Jordan-Wigner transformation enables the 25 2D nodes for $n=450$ qubits to compute with a high degree of parallelism, especially in the groups with lowest Pauli string weight.
\begin{figure}[!b]
  \centering
  \includegraphics[width=0.99\linewidth]{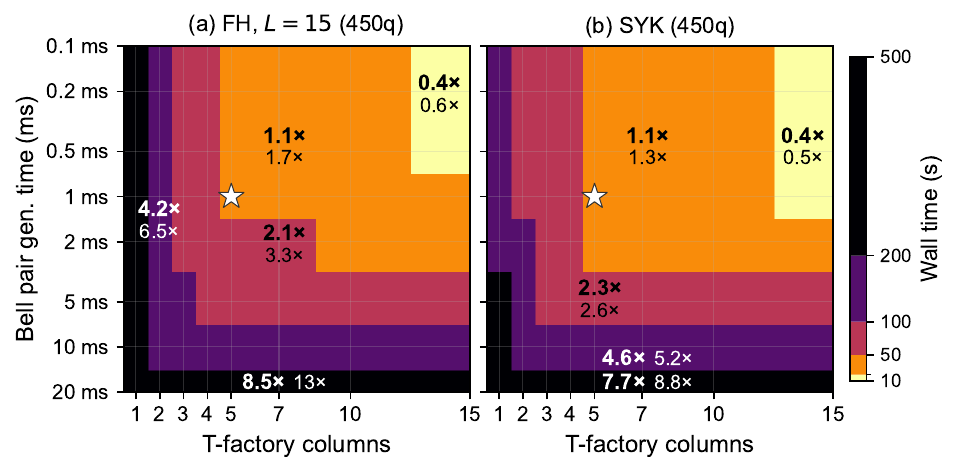}
  \caption{\textbf{Communication time and T factory production} Heatmap shows wall clock time as a function of Bell generation rate and T-factory columns. Star indicates the 1D 6L5T device. Bold represents runtime-ratios against a monolithic version of this device. Non-bold represents runtime-ratios against the 20L5T monolithic device.}
  \label{fig:heatmaps}
\end{figure}
Importantly, we see how compute time is concentrated in the groups with the highest weight and most number of non local hops required. Therefore, even though small groups can be executed quickly and locally on the 2D device, the time to compute these is negligible compared to the larger groups where the 1D device is competitive. This shows how the 1D device is able to efficiently compute the strings, which contribute most to runtime, even given its lower physical costs. We analyse this further in the next section.

%% file: network.tex
\section{Network and Magic relationship}

\textbf{Observation 5: Additional T factories provide little benefit in a slow storage access regime.}
\begin{figure}[htbp!]
  \centering
    \includegraphics[width=0.99\linewidth]{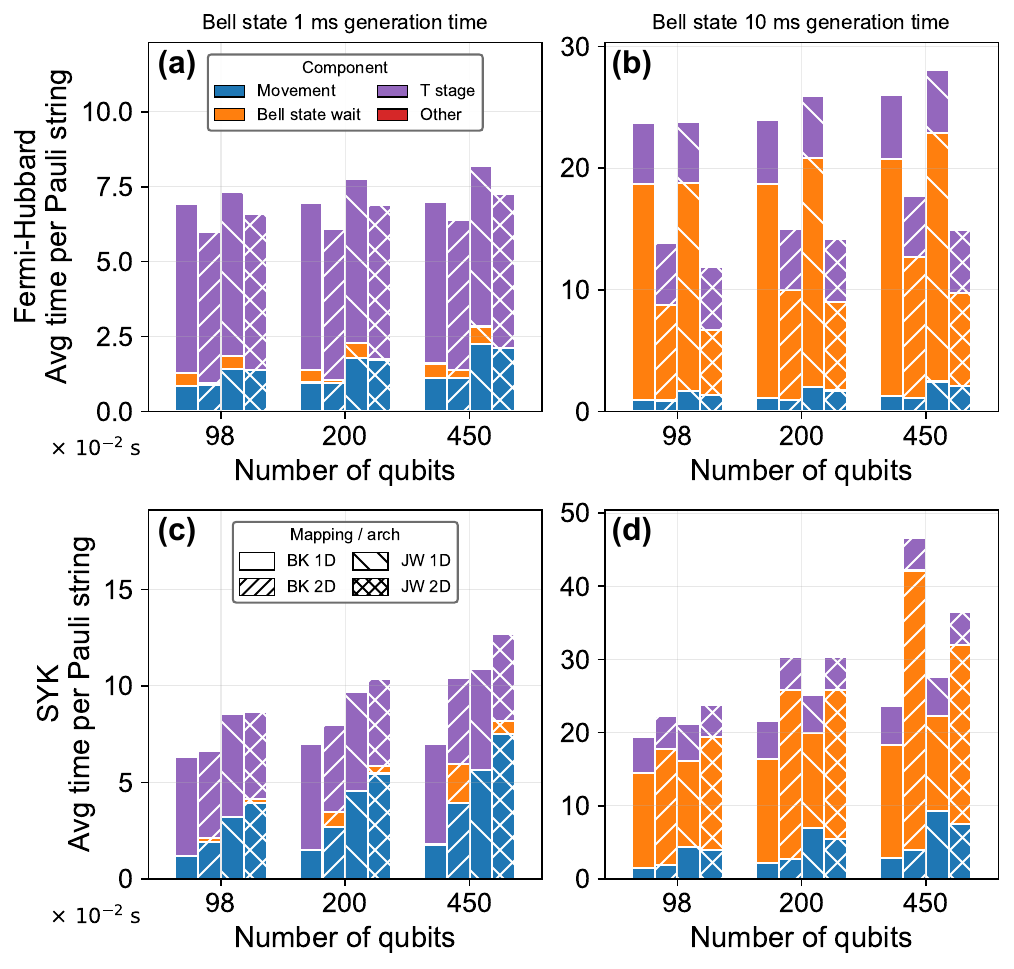}
  \caption{\textbf{Breakdown of execution time per Pauli string.} \textbf{(a-d)} Average time per Pauli string with identified blockers for different workloads, Bell state speeds, number of qubits and encoding.}
  \label{fig:bars}
\end{figure}

The heterogeneous architecture separates computation and storage into distinct subsystems. 
Consequently, overall performance is determined by the balance between the rate at which the magic core can supply non-Clifford resources and the rate at which the storage lanes can deliver parity information.
If T states are produced rapidly, the storage system must be able to compute and communicate parities at a comparable rate in order to keep the compute node occupied. 
Conversely, if T-state production is slow, improvements in communication bandwidth provide little benefit because execution becomes magic-limited.

This behavior is illustrated in Fig.\ref{fig:heatmaps}(a-b), which shows execution time as a function of both Bell pair-generation rate and T-factory count. 
When Bell pair generation is fast enough, increasing the number of T-factories leads to approximately proportional improvements in execution time, indicating that the architecture is primarily compute-limited.
However, as the Bell pair generation becomes slower, communication increasingly dominates execution. 
For example, at a Bell pair-generation time of 10 ms, little performance benefit is obtained by increasing the number of T factories beyond three, as the storage system is no longer able to supply parity information quickly enough to utilize the additional non-Clifford resources.

To better understand these trends, Fig.~\ref{fig:bars}(a-d) illustrates the average execution time of a Pauli string into its constituent components. 
At fast Bell-generation rates (1 ms), execution time is dominated by T-state generation, while communication overhead contributes only a small fraction of the total runtime. 
In contrast, at slower Bell state generation rates (10 ms), communication delays become the majority of the execution time. This effect is comparatively less pronounced for the homogeneous distributed architecture in the Fermi-Hubbard simulation, due to its ability to perform rotations locally for less distributed strings.

%% file: para.tex
\section{Parallelism}
\label{sec:para}

\textbf{Observation 6: Quantum workload structure determines usable parallelism.}

\textbf{Observation 7: For overlapping non-local workloads, intra-string parallelism is more valuable than additional parallelism across Pauli-strings.}

\begin{figure}[htbp!]
  \centering
  \includegraphics[width=0.99\linewidth]{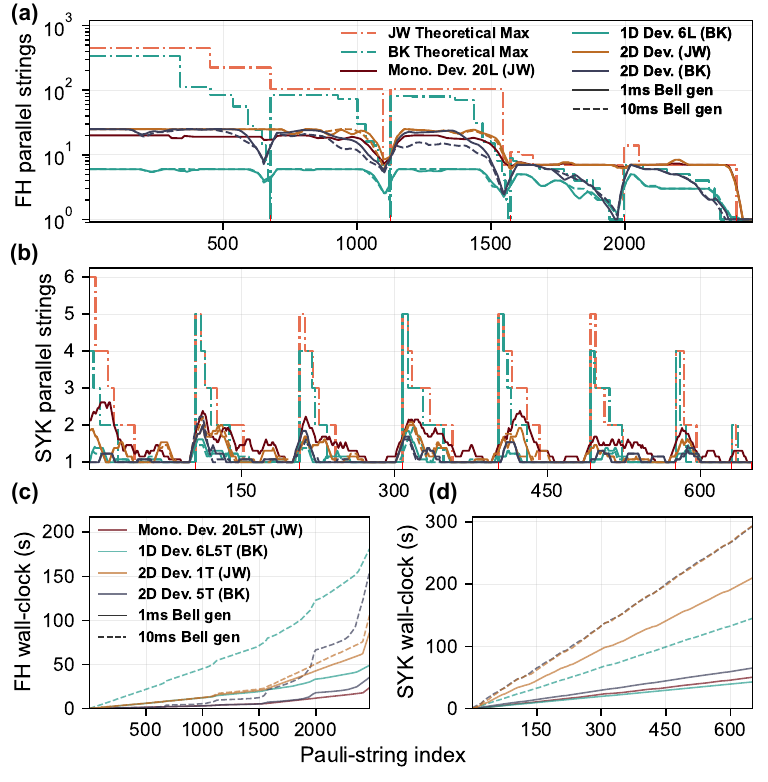}
  \caption{\textbf{Parallelizable Pauli strings and wall clock time.} \textbf{(a, b)} Graphs of the theoretical max for different encodings and actually achieved for combinations of encodings, architectures, and workloads (\textbf{a}: Fermi-Hubbard \textbf{b}: SYK), number of simultaneously executable Pauli strings throughout execution of a single Trotter step. \textbf{(c, d)} Wall clock time against number of executed Pauli strings for different architectures and workloads. Dashed lines indicate 10ms Bell pair generation speed rather than 1ms. }
  \label{fig:para}
\end{figure}

Fig.~\ref{fig:para}(a-b) depicts the degree of available and achieved parallelism for the different workloads, architectures and encodings we study. Focusing on (a), we know that at the start of the simulation the average Pauli string weight of commuting groups is small (see Fig.\ref{fig:storage}(c)). This means the theoretical maximum attainable parallelism, limited by commutativity and disjointness of the strings, is high. As the average Pauli weight increases, this decreases. We see that JW consistently has a higher degree of parallelism than BK, as expected. However, this metric becomes matched with BK later in the simulation. At earlier points, a hugely parallel device would be needed to take advantage of the theoretical limit. 

In Fig.~\ref{fig:para}(c) we see that in the regime at the start of the simulation the homogeneous architecture and monolithic baseline both initially (up to about the $1750^{th}$ Pauli string) have an increase in wall clock time with a lower gradient compared to the 1D device. The larger number of magic sites in these architectures therefore translates directly to an increased performance in this regime compared to the heterogeneous architecture. 

However, beyond this their advantage largely disappears. The remaining operators are dominated by the larger Pauli strings, whose supports can overlap extensively.
As shown by the decreasing theoretical parallelism curves in Fig.~\ref{fig:para}(a), only a small number of these strings can be executed simultaneously regardless of the amount of available hardware.
The homogeneous architecture therefore loses most of its string-level parallelism advantage.

In this regime, the heterogeneous architecture benefits from a different form of parallelism.
Rather than executing more Pauli strings simultaneously, it accelerates the execution of individual high-weight strings. 
The parity-computation circuit is distributed across multiple storage lanes, allowing the compute and uncompute stages to proceed concurrently.
Consequently, even when only a single Pauli string is being executed, the parity accumulation process can be parallelized across up to six lanes. 
The homogeneous architecture lacks an equivalent mechanism and therefore performs the parity computation largely serially along its routing path.
As a result, the largest Pauli strings, which dominate the tail of the execution profile, are completed significantly faster on the heterogeneous architecture.

This effect is accentuated in the SYK simulation because its Pauli strings are both non-local and strongly overlapping. The theoretical and achieved string-level parallelism remain close to one for much of the workload, leaving the additional execution sites of the homogeneous architecture underutilized. Performance is therefore determined primarily by the time to compute each individual parity, favoring the lane architecture’s intra-string parallelism.

%% file: disc.tex
\section{Discussion and future work}

The main concept introduced and evaluated in this work is a quantum storage system with specific access patterns and simple design. We have shown that this system can access arbitrary highly non-local Pauli strings with varying weights and non-locality faster than a homogeneous architecture (2D) that has higher connectivity. The storage lane concept enables inter- and intra-string parallelism, giving it flexibility across different workloads. The storage system is also extensible: It can adapt to workload requirements and be extended at relatively low cost, especially compared to extending a device by adding fully capable modules. As a result, the proposed architecture is competitive with homogeneous architectures, despite using significantly fewer magic factories and sites. 

The random parity access capability of the storage system enables saving in quantum simulation by making otherwise difficult, highly non-local fermionic encodings practical. This capability is likely to be useful in other quantum algorithms, too. There is scope to explore new or dynamic fermionic encodings that exploit the low cost access to non-local Pauli strings provided with this architecture. The storage system can also be modified, at some additional cost, to support tree-like or other hierarchical structures.

Further research should consider the use of qLDPC codes in place of the surface codes used in this work. Their higher encoding rate could increase storage density. Direct Pauli-based computation approaches, such as those proposed in~\cite{tourDeGross,khan2026architecting}, could also be promising within a distributed storage system such as the one introduced here. 
Improvements to T-state production and application, including analog rotations on transversal~\cite{ismail2026transversal} and extractor architectures~\cite{sethi2026injeqt}, are expected to further increase the relative importance of storage access and routing overheads.
Overall, we hope our results encourage further research into fault tolerant, extensible quantum storage systems.

%% file: methods.tex
\section{Appendix}
\label{sec:appendix}

Here we provide details about the fermionic systems and mapping discussed in this work.
\subsection{Fermionic models} \label{sec:app_models}

We consider two representative fermionic models: the
Fermi--Hubbard model~\cite{fh} and the sparse Sachdev--Ye--Kitaev~\cite{SYK}
(SYK) model.

The Fermi--Hubbard Hamiltonian on an $L\times L$ lattice
with periodic boundary conditions is
\begin{equation}
    H =
    -t
    \sum_{\sigma\in\{\uparrow,\downarrow\}}
    \sum_{\langle i,j\rangle}
    \left(
        c_{i\sigma}^{\dagger}c_{j\sigma}
        + c_{j\sigma}^{\dagger}c_{i\sigma}
    \right)
    +
    U\sum_i n_{i\uparrow}n_{i\downarrow},
\end{equation}
where $c_{i\sigma}^{\dagger}$ and $c_{i\sigma}$ are,
respectively, the fermionic creation and annihilation operators
for spin $\sigma$ at lattice site $i <L^{2}$ with $N=2L^2$, and
$n_{i\sigma}=c_{i\sigma}^{\dagger}c_{i\sigma}$ is the
corresponding number operator. The parameter $t$ denotes the
nearest-neighbor hopping strength, $U$ is the on-site
interaction strength, and $\langle i,j\rangle$ denotes pairs
of neighboring lattice sites.

We use Qiskit~\cite{qiskit2024} Nature's
\texttt{FermiHubbardModel} to generate the Hamiltonian,
with nearest-neighbor hopping amplitudes $t_{ij}=1$ and
on-site interaction strength $U=8$. Each lattice site contains
two fermionic modes, corresponding to spin up and spin down,
and is therefore represented by two qubits after
fermion-to-qubit mapping. An $L\times L$ lattice consequently
requires $2L^2$ qubits.

We consider a sparse SYK Hamiltonian as defined in~\cite{maskara2025fast}. It is defined on $N$ complex
fermionic modes, or equivalently $2N$ Majorana modes. To
match the Fermi--Hubbard qubit counts, the main sparse-SYK
calculations use $N=2L^2$ for the same nominal lattice
parameter $L$. Quartets of Majorana modes are included
independently with probability $p ={d}/{\binom{2N-1}{3}},$
where $d=3$ is the mean-degree parameter. For each included
quartet $I=(i,j,k,l)$, the coupling is sampled from a unit
normal distribution and normalized by $1/\sqrt{2dN}$, giving
{
\setlength{\abovedisplayskip}{4pt}
\setlength{\belowdisplayskip}{4pt}
\begin{equation}
    H_{\mathrm{SYK}}=
    \frac{1}{\sqrt{2dN}}
    \sum_I X_I J_I C_I ,
\end{equation}
}
where $X_I$ is a Bernoulli inclusion variable, $J_I$ is a
Gaussian coupling coefficient, $C_I$ is the
appropriately phased product of the four Majorana operators
associated with quartet $I$.

\subsection{Fermion-to-qubit transformations}\label{sec:ferm_qub_trans}

We use the Jordan--Wigner (JW) and Bravyi--Kitaev (BK)
transformations~\cite{jordan1928paulische, bravyi2002fermionic} to map the fermionic creation and annihilation
operators to qubit operators. In both cases, a fermionic
Hamiltonian is transformed into a weighted sum of Pauli
strings.

Under the Jordan--Wigner transformation, the creation and
annihilation operators for fermionic mode $i$ are mapped as
\begin{equation}
    c_i^\dagger
    \mapsto
    \frac{\left(X_i-iY_i\right)}{2} \scalebox{0.7}{$\prod$}_{j=0}^{i-1} Z_j
    ,\text{and }
    c_i
    \mapsto
    \frac{\left(X_i+iY_i\right)}{2}\scalebox{0.7}{$\prod$}_{j=0}^{i-1} Z_j
    .
\end{equation}

The string of $Z$ operators records the parity of all
fermionic modes preceding mode $i$, ensuring that the mapped
operators satisfy the fermionic anticommutation relations.
Although this mapping produces a highly regular support
structure, the length of the parity string grows linearly with
the mode index. The maximum Pauli weight therefore scales as
$O(N)$ for a system containing $N$ fermionic modes.

In the Bravyi--Kitaev transformation, occupation and parity
information are distributed across qubits using a binary-tree
encoding. Following the convention implemented by Qiskit
Nature, each fermionic mode $i$ is associated with an update
set $U(i)$, a parity set $P(i)$, and a flip set $F(i)$. We
additionally define the remainder set $R(i)=P(i)\setminus F(i).$

The two Pauli components used to represent the creation and
annihilation operators are
\begin{equation}
    A_i =
    X_i
    \left(\scalebox{0.7}{$\prod$}_{k\in U(i)}X_k\right)
    \left(\scalebox{0.7}{$\prod$}_{k\in P(i)}Z_k\right),
\end{equation}
and
\begin{equation}
    B_i =
    Y_i
    \left(\scalebox{0.7}{$\prod$}_{k\in U(i)}X_k\right)
    \left(\scalebox{0.7}{$\prod$}_{k\in R(i)}Z_k\right).
\end{equation}
The fermionic operators are then mapped according to
\begin{equation}
    c_i^\dagger
    \mapsto
    \frac{1}{2}\left(A_i-iB_i\right),
    \qquad
    c_i
    \mapsto
    \frac{1}{2}\left(A_i+iB_i\right).
\end{equation}
The update, parity, and flip sets are obtained from the
binary-tree structure and contain at most $O(\log N)$
qubits. Consequently, the resulting Pauli operators have
weight $O(\log N)$ rather than the $O(N)$ weight produced
by the Jordan--Wigner transformation. This lower weight is
obtained at the cost of a less geometrically local and less
regular support structure, because the qubits associated with
a single fermionic mode may be distributed across several
parts of the encoding tree.

\subsection{Code distance and physical-resource estimates}\label{sec:cdestimates}

Code distances are estimated from the total number of logical operations required by the Pauli-string workload.
Let $W$ be the sum of all Pauli-string weights, $M$ the number of non-identity Pauli strings, and $N_{\mathrm{step}}$ the assumed number of Trotter steps. 
The total logical-operation count is approximated as $N_{\mathrm{op}} = \left(2W + 30M\right)N_{\mathrm{step}}$ where the factor $2W$ accounts for parity computation and uncomputation, and the term $30M$ accounts for the fixed T-state cost assigned to each Pauli rotation. 
Unless otherwise stated, we assume $N_{\mathrm{step}}=10$.

To ensure that the total probability of success with respect to physical error over the full simulation remains above a reasonable $90\%$, we assign a target logical error rate per operation of $p_L = 1 - 0.9^{1/N_{\mathrm{op}}}.$

The required code distance is then estimated using the phenomenological fit $d =\left\lceil d_0 + \frac{2\log(p_L/p_0)}{\log(r)} \right\rceil$ with $d_0=11$, $p_0=5\times10^{-7}$, the physical two-qubit error rate $10^{-3}$, threshold $7\times10^{-3}$, and $r=10^{-3}/(7\times10^{-3})$.